# Atomic-environment-dependent thickness of ferroelastic domain walls

Mingqiang Li[1,2], Xiaomei Li[3], Yuehui Li[1], Heng-Jui Liu[4], Ying-Hao Chu[4,5], and Peng Gao[1,6]*

[1]Electron microscopy laboratory, School of Physics and International Center for Quantum Materials, Peking University, Beijing 100871, China

[2]Academy for Advanced Interdisciplinary Studies, Peking University, Beijing 100871, China

[3]Beijing National Laboratory for Condensed Matter Physics and Institute of Physics, Chinese Academy of Sciences, Beijing 100190, China

[4]Department of Materials Science and Engineering, National Chung Hsing University, Taichung 40227, Taiwan, ROC

[5]Institute of Physics, Academia Sinica, Taipei 11529, Taiwan, ROC

[6]Collaborative Innovation Centre of Quantum Matter, Beijing 100871, China.

E-mail address: p-gao@pku.edu.cn

ABSTRACT

Domain walls are of increasing interest in ferroelectrics because of their unique properties and potential applications in future nanoelectronics. However, the thickness of ferroelastic domain walls remains elusive due to the challenges in experimental characterization. Here, we determine the atomic structure of ferroelastic domain walls and precisely measure the polarization and domain wall thickness at picometer scale using annular bright field imaging in an aberration-corrected scanning transmission electron microscope. We find that the domain wall thickness in $PbZr_{0.2}Ti_{0.8}O_3$ and $PbTiO_3$ thin films is typically about one-unit cell, across which the oxygen octahedron distortion behavior is in excellent agreement with first principles calculations. Remarkably, wider domain walls about two-unit cells in thickness are also observed for those domains walls are coupled with dislocations and underwent a compressive strain. These results suggest that the thickness of ferroelastic domain walls highly depends on their atomic environments. This study can help us to understand the past debatable experimental results and provide further insights into control of domain walls via strain engineering for their possible applications in nanoelectronics.

## 1. Introduction

Ferroelastic domain walls possessing strong coupling of polarization and strain have attracted intense research interest for their influence on the response of ferroeletrics and increasingly being considered as functional elements in nanoelectronics due to their tiny size and unique properties.[1-6] For example, it has been reported that the presence of ferroelastic domain walls in $PbZr_{0.2}Ti_{0.8}O_3$ (PZT) films can greatly enhance the piezoelectric and dielectric response.[7-10] Ferroelastic domain walls that exhibit metallic-type conductivity are promising for future nanoelectronics applications due to their mobility and nanoscale thickness.[11-13] Furthermore, ferromagnetism was predicted at ferroelastic domain walls due to the accumulation of oxygen vacancies.[14] Magnetoelectric coupling could also be effectively controlled by ferroelastic domain switching.[15]

Understanding the microstructure of domain walls is prerequisite for their practical applications since their properties are largely governed by atomic structure.[2, 16, 17] Particularly, the thickness of ferroelastic domain wall defined by the charge and lattice order parameters can influence the dielectric and piezoelectric responses.[18, 19] as well as the mobility of domain wall which is critical in polarization switching.[2, 18, 20] Many efforts have been devoted to explore the atomic structure of the domain walls both in experiment and in theory for decades (Table S1). First-principles calculations of 90° domain walls in $PbTiO_3$ indicated a sharp change in polarization direction occurring within a transition region of 1.3 nm.[21, 22] Using Landau–Ginzburg–Devonshire theory, the width of ferroelastic domain wall was reported to be 5 nm in $PbTiO_3$.[23] Further, less than 1 nm in thickness has also been expected for the 90° domain wall in tetragonal $BaTiO_3$.[24] On the other hand, experimentally measured thickness of 90° domain wall seem to be overestimated but gradually approach the actual value with the developments of advanced characterization instruments. The width of 90° domain walls in tetragonal $BaTiO_3$ was measured to be 5-15 nm by transmission electron microscopy (TEM).[25] Later, 1.0±0.3 nm and 1.5±0.3 nm were reported in $PbTiO_3$.[26, 27] More recently, by using atomic resolution scanning transmission electron microscopy (STEM) the atomic structure of ferroelastic domain walls was studied and the width has been estimated about 5 unit cells (~2nm).[17] Nevertheless, the thickness of ferroelastic domain wall is still controversial due to the lacking of information on oxygen octahedron distortion behavior across the domain wall from previous experiments. Therefore, a precisely quantitative determination of atomic structure of ferroelastic domain walls is necessary for further exploration and domain-wall application.

In this paper, we precisely measure the atomic structure and thickness of ferroelastic domain walls in PZT and $PbTiO_3$ thin films by simultaneously determining the oxygen and cations at the domain wall with annular bright field (ABF) imaging in aberration-corrected STEM [28, 29]. The distortion of oxygen octahedron across the domain wall induces a polarization perpendicular to the domain wall, which is in excellent agreement with the previous first principles calculations. The typical width of 90° domain wall is found to be 1 unit cell in PZT films. However, a broader domain wall ~2 unit cells is also found at the same ferroelastic domain. The elastic coupling between the compressive strain of a misfit dislocation and the domain wall accounts for the widening of ferroelastic domain wall. The microstructure-dependent domain wall width might clarify the inconsistent experimental data in previous studies and also provides us strategies to control the domain wall thickness by using misfit dislocations or substrate engineering.

## 2. Experimental procedures

### *2.1. Preparation of PZT and $PbTiO_3$ films*

Tetragonal PZT films oriented in the [001] direction were grown on (001) $SrTiO_3$ substrates and the $PbTiO_3$ thin films were grown on single crystal (110)o $DyScO_3$ substrates buffered with ~50nm $SrRuO_3$ electrodes using pulsed laser deposition method. A KrF excimer laser ($\lambda$ = 248 nm) was focused on the targets with an energy density of ~ 2.5 mJ/cm$^2$ and repetition rate of 10 Hz. During the growth of the $SrRuO_3$, the temperature and the oxygen pressure were kept at 700 oC and 100 mTorr, respectively. To avoid the easy evaporation of the lead in $PbTiO_3$ thin film at high temperature, the target with 5% excess lead was used and the temperature for the subsequent growth of $PbTiO_3$ thin film was decreased to 650 oC at the same oxygen atmosphere. After the deposition process, an in situ post annealing process with the same temperature and high oxygen pressure environment of ~ 300 Torr for 30 minutes was adopted to effectively eliminate the oxygen vacancies of the sample. Finally, the sample was slowly cooled down to room temperature with a ramp rate of 5 °C/min.

### *2.2. TEM sample preparation and TEM test*

Cross-sectional TEM specimens were thinned less than 30 μm by mechanical polishing and followed by argon ion milling in a Precision Ion Polishing System 691 (Gatan). Ion milling procedure consists of two steps. In the first stage of coarse milling, the guns were at 4 keV with angles 5 ° and -5 °. In the following condition, the guns were set at 1 keV for 5 minutes with angles of 3.5 ° and -3.5 °, and further lowered to 0.1 keV for 2 minutes for final surface cleaning. ABF images were recorded at 300 kV in JEM ARM300CF (JEOL Ltd.). The convergence semi-angle for imaging is 24 mrad, collection semi-angles snap is 12 to 24 mrad for ABF imaging. High-angle annular dark field (HAADF) images used in this work were obtained from probe Cs-corrected FEI Titan 60-300 (Titan Themis in Electron Microscopy Laboratory of Peking University) operating at 300 kV.

*2.3. STEM image analysis*

Atom positions are determined by simultaneously fitting with two-dimensional Gaussian peaks to an a priori perovskite unit cell using a MatLab code. The $TiO_2$ plane between the first SrO and PbO planes is defined as the interface and labelled as the #0 atomic layer. The bond-length for atomic columns in the #*n* layer is the distance between the #*n* and #(*n+1*) layer. Taking the oxygen sublattice as the reference (i.e. $\delta u_{\mathrm{O}} = 0$), the displacements of Pb columns respective to the neighbouring O is $\delta u_{\mathrm{Pb}} = \frac{d_1 - d_2}{2}$ and Zr/Ti respective to the neighboring O is $\delta u_{\mathrm{Zr/Ti}} = \frac{d_4 - d_3}{2}$, where $d_1$, $d_2$ are long and short Pb-O bond length along *c* direction respectively, and $d_3$, $d_4$ are short and long Zr/Ti-O bond length along *c* direction respectively in Fig. 2. The bond length $d_i$ ($i$=1, 2, 3, and 4) are calculated on the basis of the fitted atomic positions. The values of Born effective charges for tetragonal $PbTiO_3$ were calculated from *ab initio* theory[30]. Geometric phase analysis (GPA) was performed using a free FRWRtools plugin for Digital Micrograph based on the original work by Hytch et al.[31].

## 3. Results

### *3.1. Atomic structure of ferroelastic domain walls*

Fig. 1 shows an atomically resolved ABF image of a ferroelastic domain wall in PZT thin film grown on a (001) $SrTiO_3$ substrate. Both the cations (Pb and ZrO/TiO) and oxygen columns are visible. The 90° domain walls is formed on the (101) plane to satisfy the requirement of minimization of the electrostatic and elastic energies.[21, 32] The arrows indicate the head-to-tail polarization configuration across a typical ferroelastic domain wall (DW (A)). The inset schematic shows the octahedral distortion behavior across the ferroelastic domain wall with respect to Pb sublattice.

To precisely extract the domain wall thickness, the atomic structure of domain wall is quantitatively measured by analyzing the oxygen octahedron distortion and polarization distribution from the ABF image. Fig. 2a shows the projection of tetragonal PZT along the [010] axis. The bond length of Pb-O and Ti-O are marked with $d_1$, $d_2$, $d_3$ and $d_4$ respectively. The Pb-O bond length map along *a* axis of DW (A) is shown in Fig. 2b (*c* axis, Fig. S1). Both maps show atomically sharp change across the ferroelastic domain wall. Fig. 2c shows a map of displacement between the cations (Pb and ZrO/TiO) and oxygen columns, which represents the spontaneous polarization.[33, 34] The deformation of the oxygen octahedron at the domain wall in good agreement with the theoretical model in which a polarization points towards the [101] direction at the domain wall.[16, 22] We find that the ratio of out-of-plane to in-plane lattice parameters undergoes a sharp change across the domain walls in Fig. 2d. We note that the ratio calculated from the Ti sublattice is relatively smooth across the domain wall in comparison with that of Pb sublattice probably because the Ti sublattice are more likely to shift under the strain field at the domain wall.[35] Besides, the spontaneous polarization in each unit cell can be calculated precisely based on

$$\delta P = \frac{e}{\Omega}\sum_{m=1}^{N} Z_m \, \delta u_m \qquad (1)$$

where $e$ is the elementary charge, $N$ is the number of atoms in the primitive unit cell, $\Omega$ is the volume of the unit cell, $Z_m$ is the Born effective charge,[30] and $\delta u_m$ is the first order change of the position vector of the $m$th basis atom which can be calculated from bond length of $d_1$, $d_2$, $d_3$ and $d_4$. Fig. 2e shows the polarization profiles of the PbO and $TiO_2$ planes across the domain wall. Two components of polarization match well with the theory in both parallel (*z* direction) and perpendicular (*x* direction) directions respect to the 90° domain wall.[21] The transition area with changed polarization is about 1 unit cell in thickness. Overall, both the lattice parameter and electric polarization unambiguously shows a narrow ferroelastic domain wall.

*3.2. Different domain wall thickness*

Interestingly, wider domain walls were also observed in the same thin film. Fig. 3a shows a HAADF image of PZT films. Ferroelastic domain walls (A and B) are marked by two dashed yellow lines while a yellow circle denotes a misfit dislocation. The DW (A) is discussed above. In fact, in many $Pb(Zr, Ti)O_3$ thin films, the ferroelastic domains were observed to coexist with misfit dislocations in order to adopt the strain relaxation as well as minimize the elastic energy.[4, 32, 36] Fig. 3b shows the distribution of polarization of two ferroelastic domain walls. The Pb-O bond length map along *c* axis in Fig. 3c (*a* axis, Fig. S2) shows DW (A) is very sharp as demonstrated above while the DW (B) is wider. The ratio of out-of-plane to in-plane lattice parameters in Fig. 3d indicates a sharp peak is located around DW (A) while none is observed at DW (B). Fig. 3e shows the profiled of spontaneous polarization in parallel (*z* direction) and perpendicular (*x* direction) directions respect to the 90° domain wall. Thickness of DW (B) ~2 unit cells is about twice of DW (A). The varied thickness of domain walls is likely related with the atomic environments.

*3.3. Atomic environment analysis of the domain walls*

In order to figure out the origin of the asymmetric domain wall thickness, we perform GPA of the ferroelastic domain walls and the misfit dislocation in Fig. 4. The image in Fig. 4a is rotated so that we can get the strain distribution along the domain wall direction clearly. Note that the dislocation $\boldsymbol{b}$ =1/$\boldsymbol{a}$[-100] can be decomposed into two half dislocations with Burgers vectors $\boldsymbol{b_1}$ =1/2$\boldsymbol{a}$[-101] and $\boldsymbol{b_2}$ =1/2$\boldsymbol{a}$[-10-1] (No.1 and No.2), which are parallel and perpendicular to the domain walls respectively. In Fig. 4b, the $\varepsilon_{xx}$ map illustrates the strain distribution along the domain wall direction. A tensile strain field at upper side and a compressive strain field at lower side are coupled with the DW (A) and DW (B) respectively. For the strain field $\varepsilon_{yy}$ shown in Fig. 4c, both of the domain walls undergo a similar tensile strain field due to the dislocation $\boldsymbol{b}_2$ =1/2***a***[-10-1]. Therefore, the domain wall thickness is likely related with the strain field of $\varepsilon_{xx}$ rather than $\varepsilon_{yy}$ (we will discuss details below). Furthermore, the lattice rotation map $\omega_{xy}$ in Fig. 4d shows that the lattice rotation around the dislocations is about 2.57° while in the *a*-domain it is about 1.48° with the substrate as the reference. Such coupling of the lattice rotation between dislocation and *a*-domain triggers the nucleation of the *a*-domain.[37]

## 4. Discussion

The elastic coupling between the dislocation core and the ferroelastic domain wall is responsible for the broadened domain wall. The effects of strain field on domain wall have been carefully studied in $PbTiO_3$ by theory.[16] The domain wall structure highly depends on the strain field along the wall direction and is insensitive to strain in other directions. Especially, tensile strain along the domain wall direction can result in a sharp transition while compressive strain can lead to a smooth change of the lattice parameter across the domain wall,[16] which is indeed in excellent agreement with our experimental observation, i.e., a peak of ratio at DW (A) while smooth transition at DW (B) is observed. This can be understood because DW (B) is subject to compressive strain while DW (A) suffers from tensile strain along the domain wall direction. In comparison, ferroelastic domain walls in $PbTiO_3$ films without dislocations are also studied to further confirm the effect of misfit dislocations. Quantitative analysis of the domain wall suggests that the thickness of these two domain walls is almost the same, about 1 unit cell (Fig. S3). Therefore, the broadened ferroelastic domain wall can be understood as the results of the elastic coupling between the dislocation and the domain wall.

Although many efforts have been made to reveal the structure of ferroelastic domain wall, the precisely atomic structure and thickness of ferroelastic domain wall remains controversial and elusive.[12, 16, 22, 38, 39] In our study, the typical width of ferroelastic domain wall is measured to be ~1 unit cell which is very close to the ideal case where ferroelastic domain wall is regard as [101] planar to separate two adjacent domains with different order parameters (charge and lattice).[1, 19, 22, 24, 40] Further, we also find asymmetric thickness of ferroelastic domain walls in the same domain due to the strain effect form the misfit dislocation. Although anomalously thick domain walls have also been reported for charged 90° domain walls,[17, 41] the head-to-tail configuration in this case can exclude the significant charge effect. The coexistence of varied width of domain wall in ferroelectrc thin films implies us that the atomic structure of domain wall is highly depends on the atomic environments due to their complicate coupling with interfaces, dislocations and other domain walls.[1, 42, 43]

The ferroelastic domain wall enjoy many novel functions since the intrinsic coupling between strain and polarization charge as well as the interaction with defects such as oxygen vacancies.[2, 12, 14] The thickness of ferroelastic domain wall, as one of the main characters, can be used to manipulate the properties of domain wall and response of ferroelectrics. For example, giant dielectric and piezoelectric response might appear in thin films as a result of the enormously large domain-wall contribution.[18] What is more, the mobility and other features of domain walls are also mediated by domain wall thickness which are critical in domain switching.[2, 14, 18] The varied thickness of ferroelastic domain walls suggests these properties can further be manipulated in a certain degree by defects design or strain engineering.

## 5. Conclusion

In summary, atomically resolved ABF images in aberration-corrected STEM has been quantitatively analyzed to reveal the atomic structure, polarization configuration and thickness of 90° domain walls in lead-based ferroelectric films. Across the domain wall, the distortion of oxygen octahedron is identified and the typical thickness of ferroelastic domain wall is measured to be about 1 unit cell. We find that even in the same ferroelastic domain the domain wall thicknesses are different due to the dislocation induced strain field, i.e., ~1 unit cell for the domain wall subjected to the tensile strain field and ~2 unit cells for the one under compressive strain field. The environment-dependent thickness of ferroelastic domain walls not only helps to understand the previous debatable experimental results but also provides valuable information for further exploration of domain walls and their applications.

**Acknowledgements**

P.G. acknowledges the supported from Prof. Yuichi Ikuhara in University of Tokyo for the use of electron microscope. P.G. was supported by the National Basic Research Program of China (2016YFA0300804), the National Natural Science Foundation of China (51672007, 11974023), Key Area R&D Program of Guangdong Province (2018B010109009), The Key R&D Program of Guangdong Province (2018B030327001), National Equipment Program of China (ZDYZ2015-1), and the "2011 Program" Peking-Tsinghua-IOP Collaborative Innovation Centre for Quantum Matter. H.-J.L. and Y.-H.C. were supported by the Ministry of Science and

Technology, R.O.C. (MOST 103-2119-M-009-003-MY3), and Centre for Interdisciplinary Science of National Chiao Tung University.

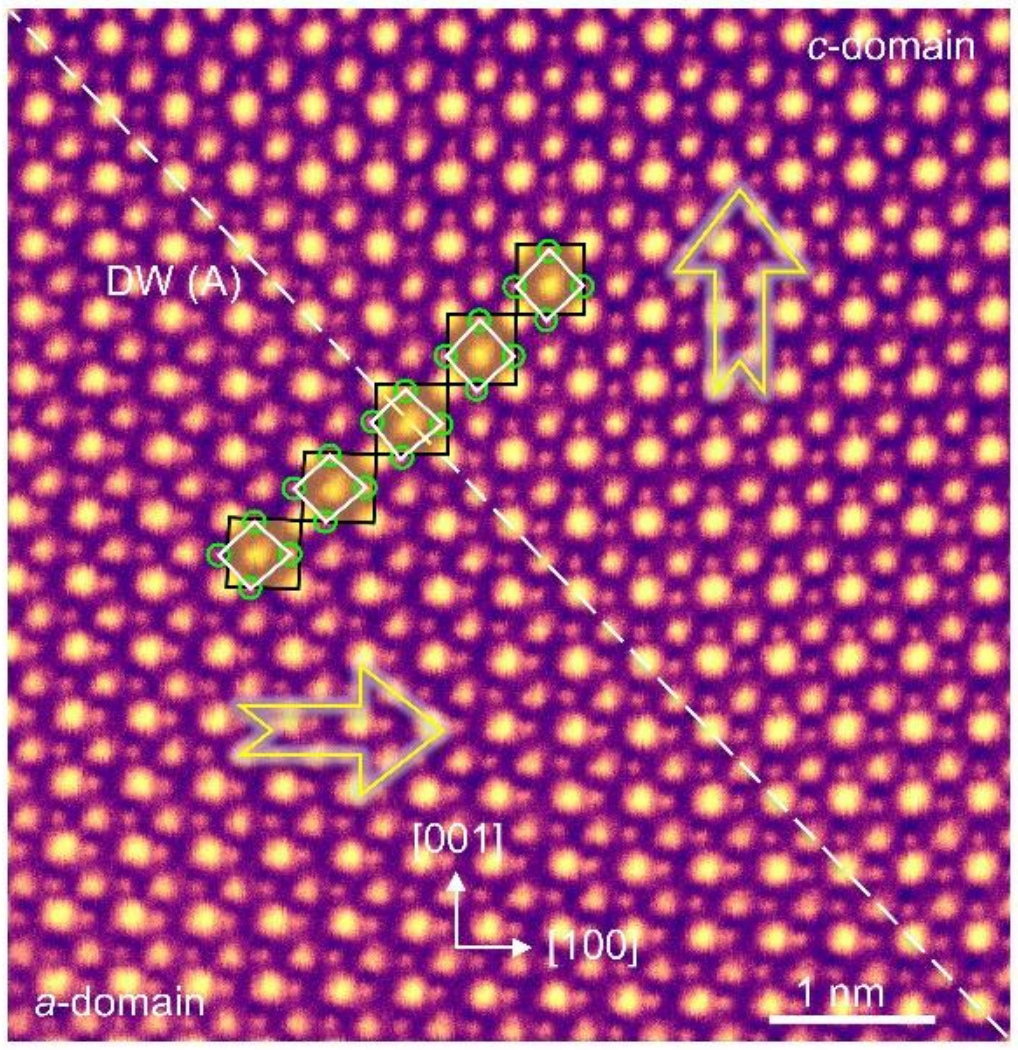

Fig. 1. A ferroelastic domain wall in a $PbZr_{0.2}Ti_{0.8}O_3$ thin film. Atomically resolved annular bright field (ABF) image of a ferroelastic domain wall in $Pb(Zr_{0.2}Ti_{0.8})O_3$ (PZT) thin film. The yellow dashed line represents the ferroelastic domain wall. The inset schematic shows the octahedral distortion behavior across the ferroelastic domain wall. The large arrows represent the polarization direction. Scale bar, 1 nm.

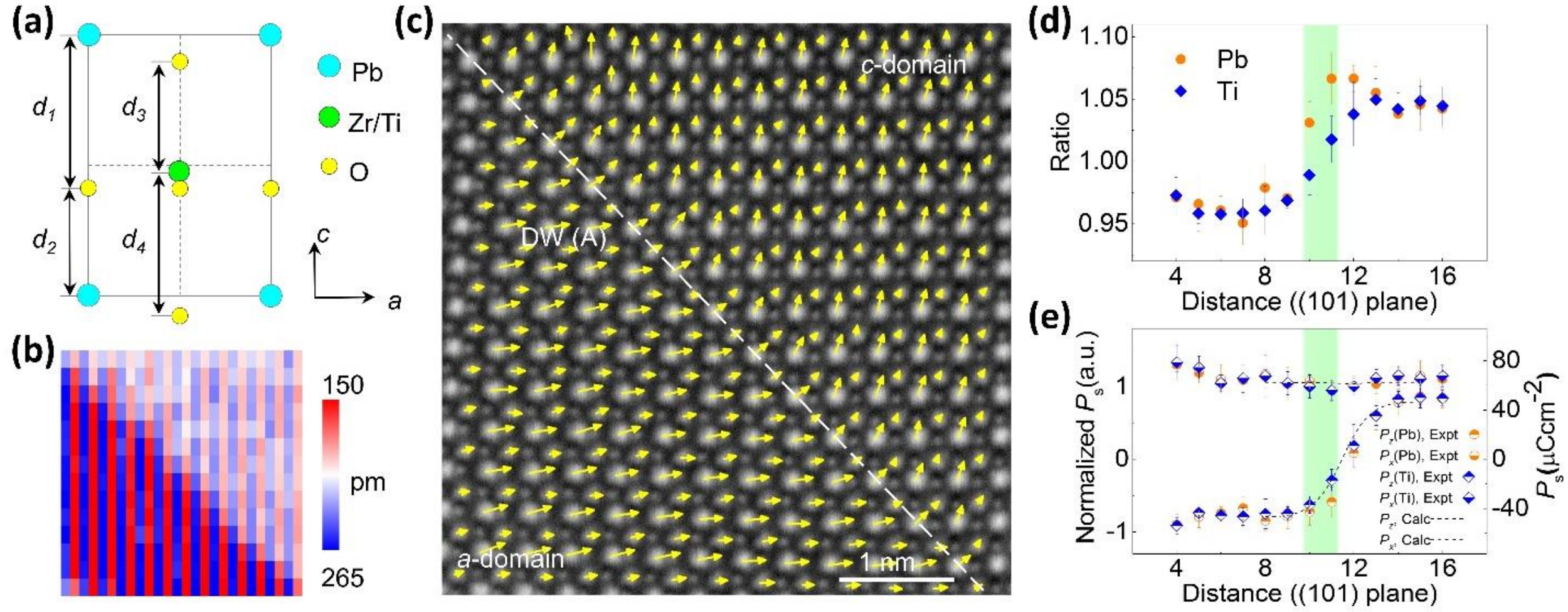

Fig. 2. The tetragonality and polarization of the ferroelastic domain wall. (a) Schematic shows the projection of tetragonal PZT along the [010] axis. (b) Pb-O bond length (both long and short length) map along *a*-axis direction. (c) The vector map of the displacements between cations and oxygen columns. (d) The ratio of out-of-plane to in-plane lattice parameters calculated from the Pb sublattice and Ti sublattice respectively. The light green color band highlights the position of the domain wall. The error bar is the standard deviation. (e) Both experimental and first principles calculations of polarization[21] in two directions: perpendicular (*x* direction) and parallel (*z* direction) to the domain wall.

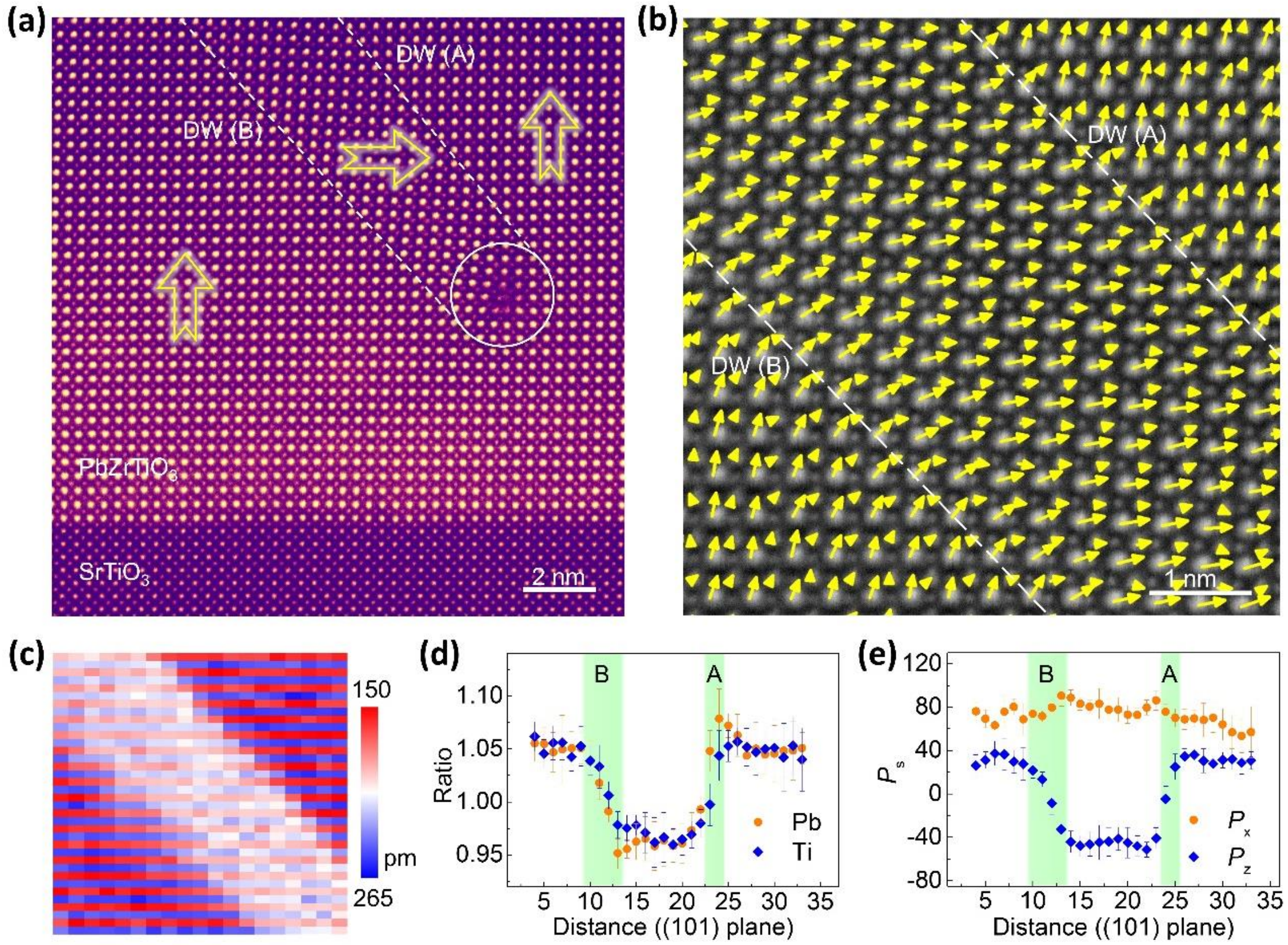

Fig. 3. Ferroelastic domain walls with different thickness. (a) Atomically resolved high angle annular dark field (HAADF) image showing ferroelastic domain and dislocations in PZT thin films. (b) The vector map of the displacement between cations and oxygen. (c) The map of Pb-O bond length (both long and short length) along vertical direction (*c* axis) showing DW (A) with sharper contrast than DW (B). (d) The ratio of out-of-plane to in-plane lattice parameters calculated from the Pb sublattice and Ti sublattice respectively. The color band highlights the different thickness of two domain walls. (e) Polarization plots for domain walls in two directions: perpendicular (*x* direction) and parallel (*z* direction) to domain walls.

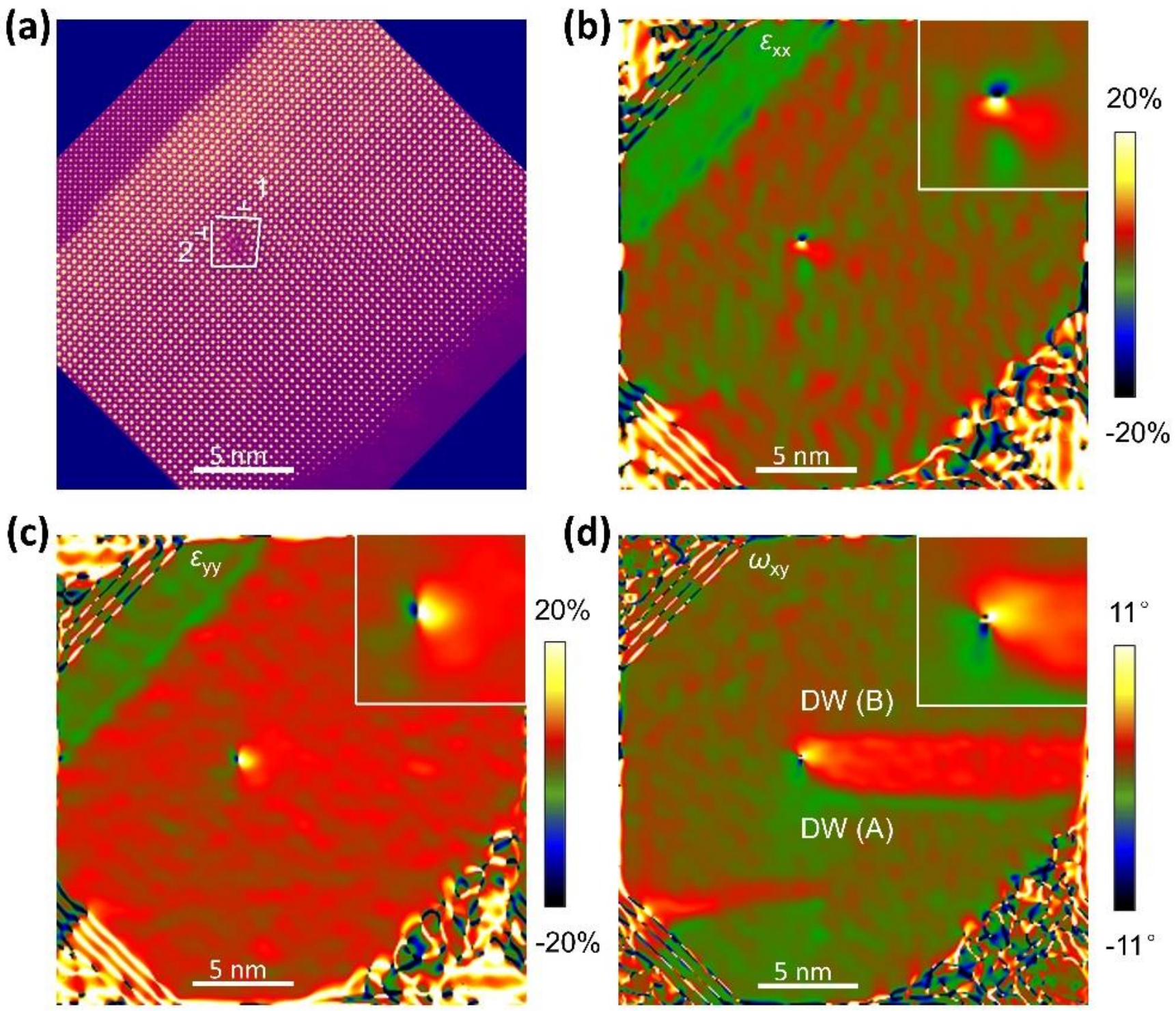

Fig. 4. GPA analysis around the *a*-domain and the misfit dislocation. (a) A rotated HAADF image of the *a*-domain and two partial dislocations with Burgers vectors ***b*** = 1/2***a***[-101] and ***b*** = 1/2***a***[-10-1]. (b) The corresponding strain maps along [-101] ($\varepsilon_{xx}$) is parallel with domain walls. (c) The corresponding strain maps along [101] ($\varepsilon_{yy}$) is perpendicular to domain walls. (d) The corresponding lattice rotation map ($\omega_{xy}$). The insets in (b)-(d) magnify strain field and lattice rotation around the dislocation core.

# Supplementary Material for

# Atomic-environment-dependent thickness of ferroelastic domain walls

Mingqiang Li[1,2], Xiaomei Li[3], Yuehui Li[1], Heng-Jui Liu[4], Ying-Hao Chu[4,5], and Peng Gao[1,6]*

[1]Electron microscopy laboratory, School of Physics and International Center for Quantum Materials, Peking University, Beijing 100871, China

[2]Academy for Advanced Interdisciplinary Studies, Peking University, Beijing 100871, China

[3]Beijing National Laboratory for Condensed Matter Physics and Institute of Physics, Chinese Academy of Sciences, Beijing 100190, China

[4]Department of Materials Science and Engineering, National Chung Hsing University, Taichung 40227, Taiwan, ROC

[5]Institute of Physics, Academia Sinica, Taipei 11529, Taiwan, ROC

[6]Collaborative Innovation Centre of Quantum Matter, Beijing 100871, China.

*To whom correspondence should be addressed.

E-mail: p-gao@pku.edu.cn (P.G.)

Table S1. Theoretical and experimental values of the 90 ° domian wall thickness in tetragonal ferroelectrics.

| 90 ° domain wall thickness | | | |
|---|---|---|---|
| Methods | Materials | Thickness (nm) | Source (Year) |
| TEM | $BaTiO_3$ | 5-15 | 1974 [1] |
| | $(Ba,Pb)TiO_3$ | | |
| TEM | $PbZr_{0.52}Ti_{0.48}O_3$ | ≤10 | 1981 [2] |
| Electron holography | $BaTiO_3$ | 1.5-2.5 | 1992 [3] |
| HRTEM | $PbZr_{0.52}Ti_{0.48}O_3$ | 1.6-4 | 1992 [4] |
| | $BaTiO_3$ | | |
| HRTEM | $PbTiO_3$ | 1 | 1995 [5] |
| HRTEM | $BaTiO_3$ | 3-4 | 1996 [6] |
| HRTEM | $BaTiO_3$ | 4-6 | 1997 [7] |
| X-ray diffraction | $BaTiO_3$ | 14 | 1999 [8] |
| HRTEM/WBTEM | $PbTiO_3$ | 1.5/2.1 | 1999 [9] |
| AFM | $PbTiO_3$ | 9-10 | 2004 [10] |
| HRTEM | $PbZr_{0.4}Ti_{0.6}O_3$ | 1.5 | 2005 [11] |
| Theory | $BaTiO_3$ | 3.6 | 2006 [12] |
| Theory | $PbTiO_3$ | 5 | 2007[13] |
| Theory | $PbTiO_3$ | 1.3 | 2008[14] |
| Theory | $BaTiO_3$ | <1 | 2010[15] |
| STEM | $PbTiO_3$ | 2 | 2014[16] |

**S1. Pb-O bond length map of DW (A)**

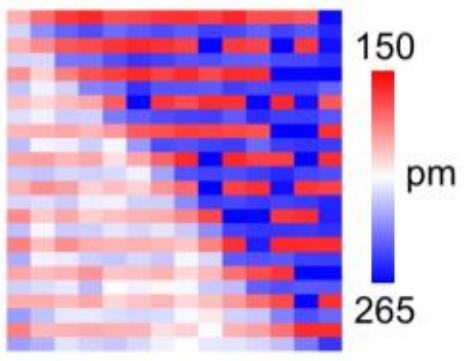

Fig. S1. Pb-O bond length (both long and short length) map of DW (A) along *c*-axis direction.

**S2. Pb-O bond length map of DW (A) and (B)**

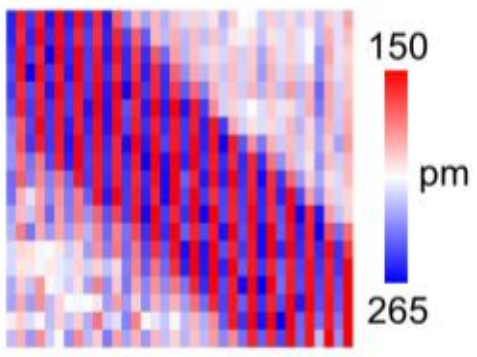

Fig. S2. Pb-O bond length (both long and short length) map of DW (A) and (B) along *a*-axis direction.

## S3. Ferroelastic domain wall in $PbTiO_3$ films

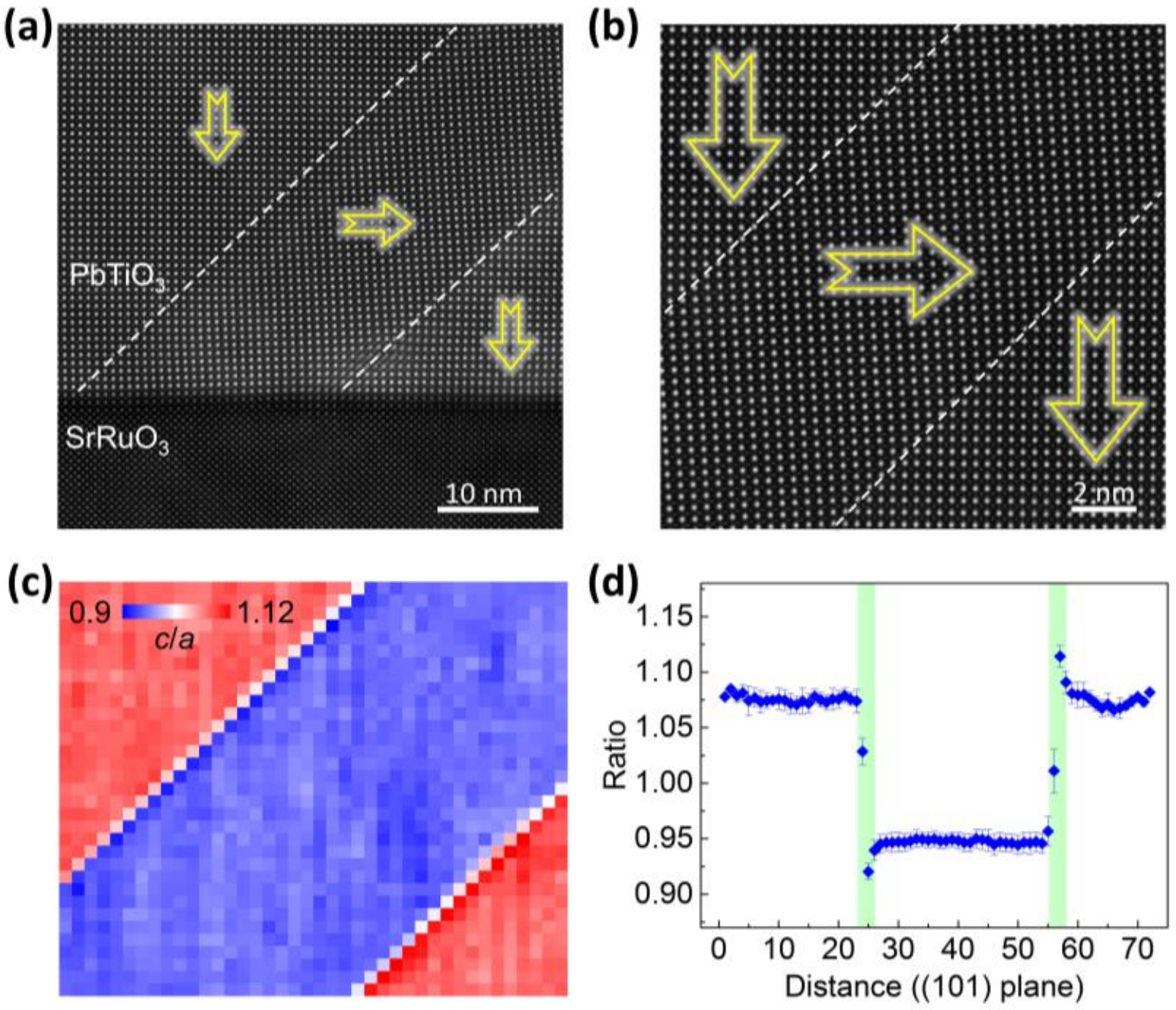

Fig. S3. Ferroelastic domain wall in $PbTiO_3$ films without misfit dislocations. (a) A HAADF image of the *a*-domain in $PbTiO_3$ thin films grown on $DyScO_3$ with 50 nm $SrRuO_3$ electrodes. No dislocation is observed at the interface. (b) Atomic resolution HAADF image of the ferroelectric domain. The dashed lines indicate two ferroelastic domain walls across which the polarization configuration is head-to-tail. (c) The *c*/*a* map of two ferroelastic domain walls. Both of the domain walls are atomically sharp. (d) The plot of the lattice parameter ratio *c*/*a*. The data points were averaged along domain walls. The error bar is the standard deviation. The color band highlights the position of two domain walls.